\documentclass[11pt]{article}
\usepackage{amssymb,amsmath,amsfonts}
\usepackage{graphicx}
\usepackage{graphics}
\usepackage{eepic,epsfig}

\begin{document}

\title{ANGULAR DISTRIBUTION OF POSITRONS IN COHERENT PAIR PRODUCTION IN
DEFORMED CRYSTALS}
\author{V. V. Parazian \\
\textit{Institute of Applied Problems in Physics,}\\
\textit{25 Nersessian Str., 0014 Yerevan, Armenia}}
\maketitle

\begin{abstract}
We investigate the angular distribution of positrons in the coherent process
electron-positron pair creation process by high-energy photons in a
periodically deformed single crystal with a complex base. The formula for
the corresponding differential cross-section is derived for an arbitrary
deformation field. The case is considered in detail when the photon enters
into the crystal at small angles with respect to a crystallographic axis.
The results of the numerical calculations are presented for ${\mathrm{SiO}}%
_{2}$ and diamond single crystals and Moliere parameterization of the
screened atomic potentials in the case of the deformation field generated by
the acoustic wave of S -type.
\end{abstract}

\bigskip


\bigskip


\bigskip

\section{Introduction}

The investigation of high-energy electromagnetic processes in crystals is of
interest not only from the viewpoint of underlying physics but also from the
viewpoint of practical applications. From the point of view of controlling
the parameters of various processes in a medium, it is of interest to
investigate the influence of external fields, such as acoustic waves,
temperature gradient etc., on the corresponding characteristics. The
considerations of concrete processes, such as diffraction radiation \cite%
{MkrtchLSh}, transition radiation \cite{LShAHMkrtch}, parametric X-radiation
\cite{MkrtchAsl}, channelling radiation \cite{MkrtchGaspar}, bremsstrahlung
by high-energy electrons \cite{Paraz41}, have shown that the external fields
can essentially change the angular-frequency characteristics of the
radiation intensities. Recently there has been broad interest to compact
crystalline undulators with periodically deformed crystallographic planes an
efficient source of high energy photons \cite{Koro97} (for a review with
more complete list of references see \cite{Koro04}).

Motivated by the fact that the basic source for the creation of positrons
for high-energy electron-positron colliders is the electron-positron pair
creation by high-energy photons (for a recent discussion see, for example,
\cite{Artru}), in \cite{MkrtchKhach} we have investigated the influence of
the hypersonic wave excited in a crystal on this process. To have an
essential influence of the acoustic wave high-frequency hypersound is
needed. Usually this type of waves is excited by high-frequency
electromagnetic field through the piezoelectric effect in crystals with a
complex base. In the paper \cite{ParazB20}, \cite{ParVard} we have
generalized the results of \cite{MkrtchKhach} for crystals with a complex
base and for acoustic waves with an arbitrary profile. For the experimental
detection of final particles in the process of coherent pair production it
is important to know their angular distribution. In the present paper the
angular distribution of positrons in the coherent pair production in
crystals is investigated in the presence of hypersonic wave. The numerical
calculations are carried out for the quartz and diamond single crystals and
for the photons of energyies 20 GeV and 100 GeV.

The paper is organized as follows. In the next section we derive the general
formula for the coherent part of the pair creation cross-section averaged on
thermal fluctuations and the conditions are specified under which the
influence of the deformation field can be considerable. In Sec. 3 the
analysis of the general formula is presented in the cases when the photon
enters into the crystal at small angles with respect to crystallographic
axes or planes and the results of the numerical calculations for the
cross-section as a function of the angle between the momenta of photon and
positron. Sec. 4 summarizes the main results of the paper.

\section{Angular Dependence of the Cross-Section}

\label{sec2:form}

By making use of the results for the bremsstrahlung derived in \cite%
{AkhBoldShul}, after the redefinition of the variables, we receive the
differential cross-section for the electron-positron pair production on an
individual atom:
\begin{eqnarray}
\frac{d^{5}\sigma _{0\pm }}{d\epsilon _{+}dq_{\Vert }d\mathbf{q}_{\bot }dy}
&=&\frac{e^{2}}{8\pi ^{4}\omega ^{2}}\frac{q_{\bot }^{2}}{q_{\Vert }^{2}}%
\left[ \frac{\omega ^{2}}{2\epsilon _{+}\epsilon _{-}}-1+4y^{2}\frac{\delta
_{\pm }}{q_{\Vert }}\left( 1-\frac{\delta _{\pm }}{q_{\Vert }}\right) \right]
\frac{\left\vert u\left( \mathbf{q}\right) \right\vert ^{2}}{\sqrt{1-y^{2}}}
\notag \\
&=&\left\vert u\left( \mathbf{q}\right) \right\vert ^{2}\sigma _{0}\left(
\mathbf{q,}y\right) ,  \label{sigindatom}
\end{eqnarray}%
where $e$ is the electron charge, $\omega $, $\epsilon _{+}$, $\epsilon _{-}$%
, are the energies of photon, positron and electron respectively (the system
of units $\hbar =c=1$ is used), $\delta _{\pm }=m_{e}^{2}\omega /\left(
2\epsilon _{+}\epsilon _{-}\right) $, $m_{e}$ is the mass of electron, $%
q_{\Vert }$ and $q_{\bot }$ are the components of the vector of momentum
transfer $\mathbf{q}$, $\mathbf{q=k-p}_{+}-\mathbf{p}_{-}$ ($\mathbf{k}$, $%
\mathbf{p}_{+}$, $\mathbf{p}_{-}$ are the momenta of photon, positron and
electron respectively), parallel and perpendicular to the direction of the
photon momentum, $u\left( \mathbf{q}\right) $ is the Fourier transform of
the atom potential. The variable $y$ is expressed in terms of the angle $%
\theta _{+}$ between the momenta $\mathbf{k}$ and $\mathbf{p}_{+}$ by the
following relation:
\begin{equation}
\left( \frac{\omega \theta _{+}}{m_{e}}\right) ^{2}=\frac{1}{\delta _{\pm }}%
\left( q_{\Vert }-\delta _{\pm }-\frac{q_{\bot }^{2}}{2\omega }+\frac{%
q_{\bot }^{2}\delta _{\pm }}{m_{e}^{2}}\right) +y\frac{2q_{\bot }}{m_{e}}%
\left( \frac{q_{\Vert }}{\delta _{\pm }}-1-\frac{q_{\bot }^{2}}{2\omega
\delta _{\pm }}\right) ^{\frac{1}{2}}.  \label{tetplyshul}
\end{equation}%
The regions of variables $q_{\parallel }$, $q_{\perp }$, $y$ in
cross-section (\ref{sigindatom}) are as follows \cite{AkhBoldShul}:
\begin{equation}
q_{\parallel }\geqslant \delta _{\pm }+\frac{q_{\perp }^{2}}{2\omega }%
,\;-1\leqslant y\leqslant 1,\;q_{\perp }\geqslant 0.  \label{paymanyq}
\end{equation}

The differential cross-section for the pair creation in a crystal can by
written in the form \cite{ParazB20}

\begin{equation}
\mathbf{\sigma }\left( \mathbf{q,}y\right) \equiv \frac{d^{5}\sigma _{0\pm }%
}{d\epsilon _{+}dq_{\parallel }d\mathbf{q}_{\perp }dy}=\left\vert
\sum_{n,j}u_{\mathbf{q}}^{\left( j\right) }e^{i\mathbf{qr}_{n}^{\left(
j\right) }}\right\vert ^{2}\sigma _{0}\left( \mathbf{q,}y\right) ,
\label{sigkrysdef}
\end{equation}%
where $\mathbf{r}_{n}^{\left( j\right) }$ is the position of an atom in the
crystal. In the discussion that follows, the collective index $n$ enumerates
the elementary cell and the subscript $j$ enumerates the atoms in a given
cell of a crystal. Here $\mathbf{q}$ is the momentum transferred to the
crystal, $\mathbf{q=k-p}_{+}-\mathbf{p}_{-}$ and the differential
cross-section in a crystal given by (\ref{sigkrysdef}), differs from the
cross-section on an isolated atom by the interference factor which is
responsible for coherent effects arising due to the periodical arrangement
of the atoms in the crystal. At non-zero temperature one has $\mathbf{r}%
_{n}^{\left( j\right) }=\mathbf{r}_{n0}^{\left( j\right) }+\mathbf{u}%
_{tn}^{\left( j\right) }$, where $\mathbf{u}_{tn}^{\left( j\right) }$ is the
displacement of $j$th atom with respect to the equilibrium positions $%
\mathbf{r}_{n0}^{\left( j\right) }$ due to the thermal vibrations. After
averaging on thermal fluctuations, the cross-section is written in the form
(see, for instance \cite{MkrtchKhach} for the case of a crystal with a
simple cell)%
\begin{equation}
\sigma \left( \mathbf{q,}y\right) =\left\{ N\sum_{j}\left\vert u_{\mathbf{q}%
}^{\left( j\right) }\right\vert ^{2}\left( 1-e^{-q^{2}\overline{%
u_{t}^{\left( j\right) 2}}}\right) +\left\vert \sum_{n,j}u_{\mathbf{q}%
}^{\left( j\right) }e^{i\mathbf{qr}_{n0}^{\left( j\right) }}e^{-\frac{1}{2}%
q^{2}\overline{u_{t}^{\left( j\right) 2}}}\right\vert ^{2}\right\} \sigma
_{0}\left( \mathbf{q,}y\right) ,  \label{sigtermik}
\end{equation}%
where $N$ is the number of cells, $\overline{u_{t}^{\left( j\right) 2}}$ is
the temperature-dependent mean-squared amplitude of the thermal vibrations
of the $j$th atom, $e^{-q^{2}\overline{u_{t}^{\left( j\right) 2}}}$ is the
corresponding Debye-Waller factor. In formula (\ref{sigtermik}) the first
term in figure braces does not depend on the direction of the vector $%
\mathbf{k}$ and determines the contribution of incoherent effects. The
contribution of coherent effects is presented by the second term. By taking
into account the formula (\ref{sigindatom}) for the cross-section on a
single atom, in the region $\omega q_{\perp }^{2}/\epsilon _{+}m_{e}^{2}\ll
1 $ the corresponding part of the cross-section in a crystal can be
presented in the form
\begin{equation}
\sigma _{c}=\frac{e^{2}}{8\pi ^{4}\omega ^{2}}\frac{q_{\perp }^{2}}{%
q_{\parallel }^{2}}\left[ \frac{\omega ^{2}}{2\epsilon _{+}\epsilon _{-}}%
-1+4y^{2}\frac{\delta _{\pm }}{q_{\parallel }}\left( 1-\frac{\delta _{\pm }}{%
q_{\parallel }}\right) \right] \frac{1}{\sqrt{1-y^{2}}}\left\vert
\sum_{n,j}u_{\mathbf{q}}^{\left( j\right) }e^{i\mathbf{qr}_{n0}^{\left(
j\right) }}e^{-\frac{1}{2}q^{2}\overline{u_{t}^{\left( j\right) 2}}%
}\right\vert ^{2}  \label{sigccrysfac}
\end{equation}

When external influences are present (for example, in the form of acoustic
waves) the positions of atoms in the crystal can be written as $\mathbf{r}%
_{n0}^{\left( j\right) }=\mathbf{r}_{ne}^{\left( j\right) }+\mathbf{u}%
_{n}^{\left( j\right) }$, where $\mathbf{r}_{ne}^{\left( j\right) }$
determines the equilibrium position of an atom in the situation without
deformation, $\mathbf{u}_{n}^{\left( j\right) }$ is the displacement of the
atom caused by the external influence. We consider deformations with the
periodical structure:
\begin{equation}
\mathbf{u}_{n}^{\left( j\right) }=\mathbf{u}_{0}f\left( \mathbf{k}_{s}%
\mathbf{r}_{ne}^{\left( j\right) }\right)  \label{deformperiod}
\end{equation}%
where $\mathbf{u}_{0}$ and $\mathbf{k}_{s}$\ are the amplitude and wave
vector corresponding to the deformation field, $f\left( x\right) $ is an
arbitrary function with the period $2\pi $, $\max $ $f\left( x\right) =1$.
In the discussion that follows, we assume that $f\left( x\right) \in
C^{\infty }\left( R\right) $. \ Note that we can disregard the dependence of
\ $\mathbf{u}_{n}^{\left( j\right) }$\ on the time coordinate for the case
of acoustic waves, as for particle energies we are interested in, the
characteristic time for the change of the deformation field is much greater
compared with the passage time of particles trough the crystal. For
deformation field given by Eq.(\ref{deformperiod}) the sum over the atoms in
Eq. (\ref{sigtermik}) can be transformed into the form%
\begin{equation}
\sum_{n}u_{\mathbf{q}}^{\left( j\right) }e^{i\mathbf{qr}_{n0}^{\left(
j\right) }}=\sum_{m=-\infty }^{\infty }F_{m}\left( \mathbf{qu}_{0}\right)
\sum_{n}u_{\mathbf{q}}^{\left( j\right) }e^{i\mathbf{q}_{m}\mathbf{r}%
_{ne}^{\left( j\right) }},  \label{sumntermikFm}
\end{equation}%
where $\mathbf{q}_{m}=\mathbf{q}+m\mathbf{k}_{s}$ and $F_{m}\left( x\right) $
is the Fourier transform of the function $e^{ixf\left( t\right) }$:%
\begin{equation}
F_{m}\left( x\right) =\frac{1}{2\pi }\int_{-\pi }^{\pi }e^{ixf\left(
t\right) -imt}dt.  \label{Fmdefin}
\end{equation}%
Below we need to have the asymptotic behavior of this function for large
values of \ $m$. For a fixed $x$ and under the assumptions for the function $%
f\left( x\right) $ given above, by making use the stationary phase method we
can see that $F_{m}\left( x\right) \sim O\left( |m|^{-\infty }\right) $ for $%
m\longrightarrow \infty $.

For a lattice with a complex cell the coordinates of the atoms can be
written as $\mathbf{r}_{ne}=\mathbf{R}_{n}+\mathbf{\rho }_{j}$, with $%
\mathbf{R}_{n}$ being the positions of the atoms for one of primitive
lattices, and $\mathbf{\rho }_{j}$ are the equilibrium positions for other
atoms inside $n$-th elementary cell with respect to $\mathbf{R}_{n}$. By
taking this into account, one obtains%
\begin{equation}
\sum_{m=-\infty }^{\infty }F_{m}\left( \mathbf{qu}_{0}\right) \sum_{j,n}u_{%
\mathbf{q}}^{\left( j\right) }e^{-\frac{1}{2}q^{2}\overline{u_{t}^{\left(
j\right) 2}}}e^{i\mathbf{q}_{m}\mathbf{r}_{ne}^{\left( j\right)
}}=\sum_{m=-\infty }^{\infty }F_{m}\left( \mathbf{qu}_{0}\right) S\left(
\mathbf{q,q}_{m}\right) \sum_{n}e^{i\mathbf{q}_{m}\mathbf{R}_{n}},
\label{summjnStructur}
\end{equation}%
where%
\begin{equation}
S\left( \mathbf{q,q}_{m}\right) =\sum_{j}u_{\mathbf{q}}^{\left( j\right)
}e^{i\mathbf{q}_{m}\mathbf{\rho }^{\left( j\right) }}e^{-\frac{1}{2}q^{2}%
\overline{u_{t}^{\left( j\right) 2}}},  \label{structurfacdef}
\end{equation}%
is the factor determined by the structure of the elementary cell. For thick
crystals the sum over cells in (\ref{summjnStructur}) can be presented as a
sum over the reciprocal lattice:%
\begin{equation}
\sum_{n}e^{i\mathbf{q}_{m}\mathbf{R}_{m}}=\frac{\left( 2\pi \right) ^{3}}{%
\Delta }\sum_{\mathbf{g}}\delta \left( \mathbf{q-g}_{m}\right) ,\mathbf{g}%
_{m}=\mathbf{g-}m\mathbf{k}_{s},  \label{sumdeltagmdef}
\end{equation}%
where $\Delta $ is the unit cell volume, and $\mathbf{g}$ is the reciprocal
lattice vector. Due to the $\delta $ -function in this formula, the
corresponding momentum conservation is written in the form%
\begin{equation}
\mathbf{k=p}_{+}+\mathbf{p}_{-}+\mathbf{g-}m\mathbf{k}_{s},
\label{momentumconserv}
\end{equation}%
where $-m\mathbf{k}_{s}$ stands for the momentum transfer to the external
field. As the main contribution into the coherent part of the cross-section
comes from the longitudinal momentum transfer of on order $\delta $ the
influence of the external excitation may by considerable if $|m|k_{s}$ is of
an order $\delta $. The corresponding condition will be specified later.
Another consequence of the $\delta $ -function in (\ref{sumdeltagmdef}) is
that the function (\ref{Fmdefin}) enters into the cross-section in the form $%
F_{m}\left( \mathbf{g}_{m}\mathbf{u}_{0}\right) $. Now it can be seen that
in the sum over $m$ in (\ref{summjnStructur}) the main contribution comes
from the terms for which $\left\vert m\mathbf{k}_{s}\mathbf{u}%
_{0}\right\vert \lesssim \left\vert \mathbf{gu}_{0}\right\vert $, or
equivalently $\left\vert m\right\vert \lesssim \lambda _{s}/a$, where $%
\lambda _{s}=2\pi /k_{s}$ is the wavelength of the external excitation, and $%
a$ is of the order of the lattice spacing. Indeed, for the terms with $%
\left\vert m\mathbf{k}_{s}\mathbf{u}_{0}\right\vert \gg \left\vert \mathbf{gu%
}_{0}\right\vert $ one has $F_{m}\left( \mathbf{g}_{m}\mathbf{u}_{0}\right)
\approx F_{m}\left( m\mathbf{k}_{s}\mathbf{u}_{0}\right) $, and the phase of
the integrand in (\ref{Fmdefin}) is equal to $m\left[ \mathbf{k}_{s}\mathbf{u%
}_{0}f\left( t\right) -t\right] $. Under the condition $|\mathbf{k}_{s}%
\mathbf{u}_{0}f^{\prime }\left( t\right) |<1$ this phase has no stationary
point and one has $F_{m}\left( m\mathbf{k}_{s}\mathbf{u}_{0}\right) =O\left(
|m|^{-\infty }\right) $, $m\longrightarrow \infty $ and the corresponding
contribution is strongly suppressed. By taking into account that for
practically important cases one has $\mathbf{k}_{s}\mathbf{u}_{0}\sim
u_{0}/\lambda _{s}\ll 1$, we see that the assumption made means that the
derivative $f^{\prime }\left( t\right) $ is not too large. In the way
similar to that used in \cite{ParazB20}, it can be seen that the square of
the modulus for the sum (\ref{sumntermikFm}) is written as%
\begin{equation}
\left\vert \sum_{n,j}u_{\mathbf{q}}^{\left( j\right) }e^{i\mathbf{qr}%
_{n0}^{\left( j\right) }}e^{-\frac{1}{2}q^{2}\overline{u_{t}^{\left(
j\right) 2}}}\right\vert ^{2}=N\frac{\left( 2\pi \right) ^{3}}{\Delta }%
\sum_{m,\mathbf{g}}\left\vert F_{m}\left( \mathbf{g}_{m}\mathbf{u}%
_{0}\right) \right\vert ^{2}\left\vert S\left( \mathbf{g}_{m},\mathbf{g}%
\right) \right\vert ^{2}.  \label{crystalfactor}
\end{equation}%
where $N$ is the number of cells.

Substituting this expression into formula (\ref{sigccrysfac}) and
integrating over the vector $\mathbf{q}$ by using the $\delta $ -function,
for the cross-section one obtains%
\begin{equation}
d\sigma =\int \sigma \left( \mathbf{q}\right) d^{3}q=N\left( d\sigma
_{n}+d\sigma _{c}\right) ,  \label{incohandcoh}
\end{equation}%
with $d\sigma _{n}$ and $d\sigma _{c}$ being the incoherent and coherent
parts of the cross-section per atom and $N_{0}$ is the number of atoms in
the crystal. The coherent part of the cross-section is determined by the
formula%
\begin{equation}
\frac{d^{2}\sigma _{\pm }^{c}}{d\epsilon _{+}dy}=\frac{e^{2}N}{\pi \omega
^{2}N_{0}\Delta }\sum_{m,\mathbf{g}}\frac{g_{m\perp }^{2}}{g_{m\parallel
}^{2}}\left[ \frac{\omega ^{2}}{2\epsilon _{+}\epsilon _{-}}-1+4y^{2}\frac{%
\delta _{\pm }}{g_{m\parallel }}\left( 1-\frac{\delta _{\pm }}{g_{m\parallel
}}\right) \right] \frac{\left\vert F_{m}\left( \mathbf{g}_{m}\mathbf{u}%
_{0}\right) \right\vert ^{2}\left\vert S\left( \mathbf{g}_{m},\mathbf{g}%
\right) \right\vert ^{2}}{\sqrt{1-y^{2}}},  \label{sigcohgeneral}
\end{equation}%
where the vector $\mathbf{g}_{m}$ is defined by relation (\ref{sumdeltagmdef}%
) and now the relation between the variables $y$ and $\theta _{+}$ is
written in the form:%
\begin{equation}
y=\frac{m_{e}}{2g_{m\perp }}\frac{\left( \omega \theta _{+}/m_{e}\right)
^{2}-1/\delta _{\pm }\left( g_{m\parallel }-\delta _{\pm }-g_{m\perp
}^{2}/\left( 2\omega \right) +g_{m\perp }^{2}\delta _{\pm }/m_{e}^{2}\right)
}{\left[ g_{m\parallel }/\delta _{\pm }-1-g_{m\perp }^{2}/\left( 2\omega
\delta _{\pm }\right) \right] ^{\frac{1}{2}}}  \label{ydefincrys}
\end{equation}

The regions of variables in cross-section (\ref{sigkrysdef}) are%
\begin{equation}
g_{m\parallel }\geq \delta _{\pm }+\frac{g_{m\perp }^{2}}{2\omega },\;-1\leq
y\leq 1,\;g_{m\perp }\geq 0.  \label{gmparalygmpercond}
\end{equation}%
For sinusoidal deformation field, $\ f\left( z\right) =\sin \left( z+\varphi
_{0}\right) $, one has the Fourier-transform%
\begin{equation}
F_{m}\left( x\right) =e^{im\varphi _{0}}J_{m}\left( x\right) ,
\label{FmBessel}
\end{equation}%
with the Bessel function $J_{m}\left( x\right) $.

The formula for the pair creation in an undeformed crystal is obtained from (%
\ref{sigcohgeneral}) taking $\mathbf{u}_{0}=0$. In this limit, the
contribution of the term with $m=0$ remains only with $F_{0}\left( 0\right)
=1$. Now we see that formula (\ref{sigcohgeneral}) differs from the formula
in an undeformed crystal by the replacement $\mathbf{g\rightarrow g}_{m}$,
and by the additional summation over $m$ with the weights $|F_{m}\left(
\mathbf{g}_{m}\mathbf{u}_{0}\right) |^{2}$. This corresponds to the presence
of an additional one-dimentional superlattice with the period $\lambda _{s}$
and the reciprocal lattice vector $m\mathbf{k}_{s}$, $m=0,\pm 1,\pm 2,...$ .
As the main contribution into the cross-section comes from the terms with $%
g_{m\parallel }\sim \delta _{\pm }$, the influence of the deformation field
may be considerable if $|mk_{s\parallel }|\gtrsim \delta _{\pm }$. Combining
this with the previous estimates, we find the following condition: $%
u_{0}/\lambda _{s}\gtrsim a/4\pi ^{2}l_{c}$. At high energies one has $%
a/l_{c}\ll 1$ and this condition can be consistent with the condition $%
u_{0}/\lambda _{s}\ll 1$.

In the presence of the deformation field the number of possibilities to
satisfy the condition $g_{m\parallel }\geq \delta _{\pm }+g_{m\perp
}^{2}/\left( 2\omega \right) $ in the summation of formula (\ref%
{sigcohgeneral}) increases due to the term $mk_{s\parallel }$ in the
expression for $g_{m\parallel }$. This leads to the appearance of additional
peaks in the angular distribution of the radiated positrons. After the
integration of (\ref{sigcohgeneral}) over $y$, due to these additional
peaks, there can be an enhancement of the cross-section of the process \cite%
{ParazB20}.

\section{Limiting Cases and numerical results}

\label{sec3:an}

In the following text, we consider the case when the photon enters into the
crystal at small angle $\theta $\ with respect to the crystallographic $z$\
-axis of the orthogonal lattice. The corresponding reciprocal lattice vector
components are $g_{i}=2\pi n_{i}/a_{i},$ $n_{i}=0,\pm 1,\pm 2,,...$, where $%
a_{i},$ $i=1,2,3,$ are the lattice constants in the corresponding
directions. For the longitudinal component we can write
\begin{equation}
g_{m\parallel }=g_{mz}\cos \theta +\left( g_{my}\cos \alpha +g_{mx}\sin
\alpha \right) \sin \theta ,  \label{gmpargen}
\end{equation}%
where $\alpha $ is the angle between the projection of the vector $\mathbf{k}
$ on the plane $(x,y)$ and axis $y$. For small angles $\theta $ the main
contribution into the cross-section comes from the summands with $g_{z}=0$.
Having made the replacement of variable $y\rightarrow \omega \theta
_{+}/m_{e}$ using the formula (\ref{ydefincrys}) from formula (\ref%
{sigcohgeneral}) one finds%
\begin{eqnarray}
\frac{d^{2}\sigma _{\pm }^{c}}{d\epsilon _{+}d\left( \omega \theta
_{+}/m_{e}\right) } &\approx &\frac{e^{2}N}{\pi \omega ^{2}N_{0}\Delta }%
\sum_{m,g_{x},g_{y}}\frac{g_{\perp }^{2}}{g_{m\parallel }^{2}}\left[ \frac{%
\omega ^{2}}{2\epsilon _{+}\epsilon _{-}}-1+4y^{2}\left( \theta _{+}\right)
\frac{\delta _{\pm }}{g_{m\parallel }}\left( 1-\frac{\delta _{\pm }}{%
g_{m\parallel }}\right) \right]  \notag \\
&&\times \frac{\left\vert F_{m}\left( \mathbf{g}_{m}\mathbf{u}_{0}\right)
\right\vert ^{2}\left\vert S\left( \mathbf{g}_{m},\mathbf{g}\right)
\right\vert ^{2}}{\sqrt{1-y^{2}\left( \theta _{+}\right) }}\frac{\omega
\theta _{+}/m_{e}}{\left( g_{\perp }/m_{e}\right) \left( g_{m\parallel
}/\delta _{\pm }-1-g_{\perp }^{2}/\left( 2\omega \delta _{\pm }\right)
\right) ^{\frac{1}{2}}},  \label{sigcasegz0}
\end{eqnarray}%
where the notation $y^{2}\left( \theta _{+}\right) $ is introduced in
accordance with:
\begin{equation}
y^{2}\left( \theta _{+}\right) =\frac{m_{e}^{2}}{4g_{\perp }^{2}}\frac{\left[
\left( \omega \theta _{+}/m_{e}\right) ^{2}-\left( 1/\delta _{\pm }\right)
\left( g_{m\parallel }-\delta _{\pm }-g_{\perp }^{2}/\left( 2\omega \right)
+g_{\perp }^{2}\delta _{\pm }/m_{e}^{2}\right) \right] ^{2}}{g_{m\parallel
}/\delta _{\pm }-1-g_{m\perp }^{2}/\left( 2\omega \delta _{\pm }\right) }.
\label{y2casegz0}
\end{equation}%
In (\ref{sigcasegz0}) $g_{\perp }^{2}=g_{x}^{2}+g_{y}^{2}$, and the
summation goes over the region $g_{m\parallel }\geq \delta _{\pm }+g_{m\perp
}^{2}/\left( 2\omega \right) ,$ \ $0\leq y^{2}\left( \theta _{+}\right) \leq
1$ with
\begin{equation}
g_{m\parallel }\approx -mk_{z}+\left( g_{mx}\sin \alpha +g_{my}\cos \alpha
\right) \theta .  \label{gmparmain}
\end{equation}%
Note that in the argument of the functions $F_{m}$ and $S$ we have $\mathbf{g%
}_{m}\approx \left( g_{x},g_{y},0\right) $.

We now assume that the photon enters into the crystal at small angle $\theta
$ with respect to the crystallographic axis $z$ and near the
crystallographic plane $\left( y,z\right) $ (the angle $\alpha $\ is small).
In this case with the change of $\delta _{\pm }$, the sum over $g_{x}$ and $%
g_{y}$ will drop sets of terms which lead to the abrupt change of the
corresponding cross-section. Two cases have to distinguish. Under the
condition $\delta _{\pm }\sim 2\pi \theta /a_{2}$,in Eq. (\ref{sigcasegz0})
for the longitudinal component, one has
\begin{equation}
g_{m\parallel }\approx -mk_{s\parallel }+\theta g_{y}\geq \delta _{\pm }+%
\frac{g_{\perp }^{2}}{2\omega }.  \label{gmpartetgy}
\end{equation}

The formula (\ref{sigcasegz0}) can be further simplified under the
assumption $\mathbf{u}_{0}\perp \mathbf{a}_{1}$. In this case, in the
argument of the function $F_{m}$, one has $\mathbf{g}_{m}\mathbf{u}%
_{0}\approx g_{y}u_{0y}$ and we obtain the formula

\begin{eqnarray}
\frac{d^{2}\sigma _{\pm }^{c}}{d\epsilon _{+}d\left( \omega \theta
_{+}/m_{e}\right) } &\approx &\frac{e^{2}N}{\pi ^{2}\omega ^{2}N_{0}\Delta }%
\sum_{m,g_{x},g_{y}}\frac{g_{\perp }^{2}}{g_{m\parallel }^{2}}\left[ \frac{%
\omega ^{2}}{2\epsilon _{+}\epsilon _{-}}-1+4y^{2}\left( \theta _{+}\right)
\frac{\delta _{\pm }}{g_{m\parallel }}\left( 1-\frac{\delta _{\pm }}{%
g_{m\parallel }}\right) \right]  \notag \\
&&\frac{\left\vert F_{m}\left( g_{y}u_{y0}\right) \right\vert ^{2}\left\vert
S\left( \mathbf{g}_{m},\mathbf{g}\right) \right\vert ^{2}}{\sqrt{%
1-y^{2}\left( \theta _{+}\right) }}\frac{\omega \theta _{+}/m_{e}}{\left(
g_{\perp }/m_{e}\right) \left[ g_{m\parallel }/\delta _{\pm }-1-g_{\perp
}^{2}/\left( 2\omega \delta _{\pm }\right) \right] ^{\frac{1}{2}}}.
\label{sigsumgxgy26}
\end{eqnarray}

In the second case, we assume that $\delta _{\pm }\sim 2\pi \theta \alpha
/a_{1}$. Now the main contribution into the sum in Eq. (\ref{sigcasegz0})
comes from terms with $g_{y}=0$ and summations remain over $m$ and $n_{1}$, $%
g_{x}=2\pi n_{1}/a_{1}$. The formula for the cross-section takes the form%
\begin{eqnarray}
\frac{d^{2}\sigma _{\pm }^{c}}{d\epsilon _{+}d\left( \omega \theta
_{+}/m_{e}\right) } &\approx &\frac{e^{2}N}{\pi ^{2}\omega ^{2}N_{0}\Delta }%
\sum_{m,n_{1}}\frac{g_{m\perp }^{2}}{g_{m\parallel }^{2}}\left[ \frac{\omega
^{2}}{2\epsilon _{+}\epsilon _{-}}-1+4y^{2}\left( \theta _{+}\right) \frac{%
\delta _{\pm }}{g_{m\parallel }}\left( 1-\frac{\delta _{\pm }}{g_{m\parallel
}}\right) \right]  \notag \\
&&\frac{\left\vert F_{m}\left( \mathbf{g}_{m}\mathbf{u}_{0}\right)
\right\vert ^{2}\left\vert S\left( \mathbf{g}_{m},\mathbf{g}\right)
\right\vert ^{2}}{\sqrt{1-y^{2}\left( \theta _{+}\right) }}\frac{\omega
\theta _{+}/m_{e}}{\left( g_{m\perp }/m_{e}\right) \left[ g_{m\parallel
}/\delta _{\pm }-1-g_{m\perp }^{2}/\left( 2\omega \delta _{\pm }\right) %
\right] ^{\frac{1}{2}}},  \label{sigsumg127}
\end{eqnarray}%
where
\begin{equation}
g_{m\parallel }\approx -mk_{z}+g_{x}\psi ,\quad \psi =\alpha \theta ,
\label{gmparpsi}
\end{equation}%
and the summation goes over the values $m$ and $n_{1}$ satisfying the
condition $g_{m\parallel }\geq \delta _{\pm }+g_{x}^{2}/\left( 2\omega
\right) $.

We have carried out numerical calculations for the pair creation
cross-section for various values of parameters in the case of ${\mathrm{SiO}}%
_{2}$ single crystal at zero temperature. To deal with an orthogonal
lattice, we choose as an elementary cell the cell including 6 atoms of
silicon and 12 atoms of oxygen (Shrauf elementary cell \cite{DanaDana}). For
this choice the $y$ and $z$ axes of the orthogonal coordinate system $\left(
x,y,z\right) $ coincide with the standard $Y$ and $Z$ -axes of the quartz
crystal, whereas the angle between the axes $x$ and $X$ is equal to $\pi /6$%
. For the potentials of atoms we take Moliere parametrization with%
\begin{equation}
u_{\mathbf{q}}^{\left( j\right) }=\sum_{i=1}^{3}\frac{4\pi Z_{j}e^{2}\alpha
_{i}}{q^{2}+\left( \chi _{i}/R_{j}\right) ^{2}}  \label{Molierpot}
\end{equation}%
where $\alpha _{i}=\left\{ 0.1,0.55,0.35\right\} ,$ $\chi _{i}=\left\{
6.0,1.2,0.3\right\} ,$ $R_{j}$ is the screening radius for the $j$-th atom
in the elementary cell.

The calculations are carried out for the sinusoidal transversal acoustic
wave of the S-type (the corresponding parameters can be found in Ref. \cite%
{Shaskol}) for which the vector of the amplitude of the displacement is
directed along $X$ direction of quartz single crystal, $\mathbf{u}%
_{0}=\left( u_{0},0,0\right) $, and the velocity is $4.687\cdot 10^{5}cm/$%
sec. The vector determining the direction of the hypersound propagation lies
in the plane $YZ$ and has the angle with the axis $Z$ equal to $0.295$ rad.
As the axis $z$ we choose the axis $Z$ of the quartz crystal. The
corresponding function $F\left( x\right) $ is determined by formula (\ref%
{Fmdefin}). In order to illustrate the dependence of the results on the type
of crystal we also present the numerical data for the diamond monocrystal.

\begin{figure}[tbph]
\begin{center}
\begin{tabular}{ccc}
\epsfig{figure=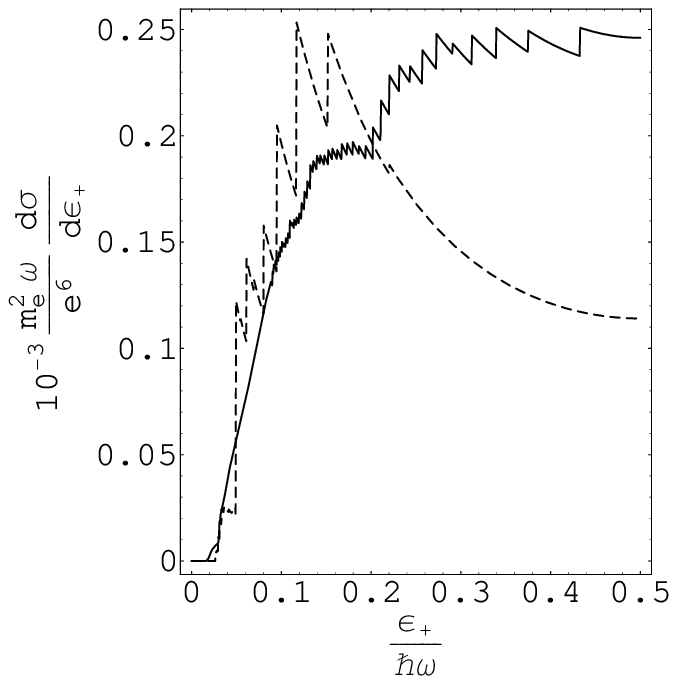,width=6cm,height=6cm} & \hspace*{0.5cm} & %
\epsfig{figure=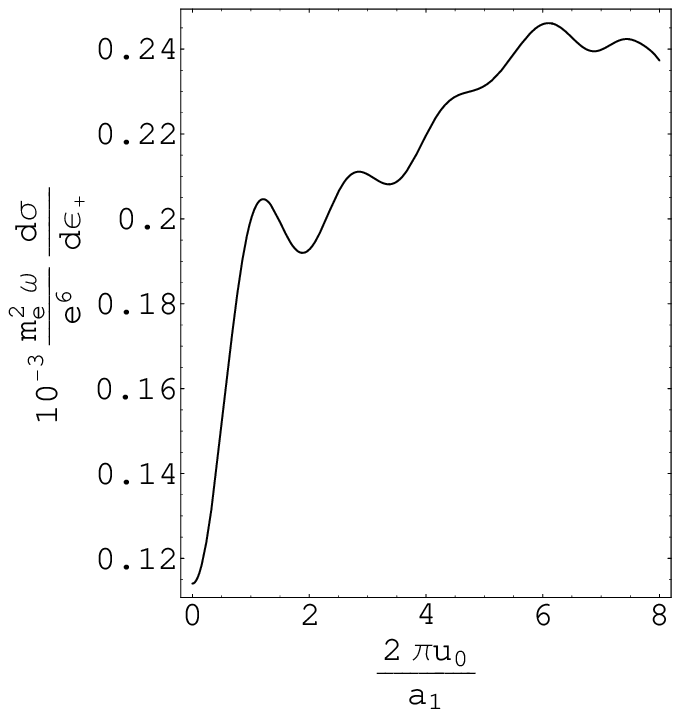,width=6cm,height=6cm}%
\end{tabular}%
\end{center}
\caption{Coherent pair creation cross-section, $10^{-3}(m_{e}^{2}\protect%
\omega /e^{6})d\protect\sigma _{\pm }^{c}/d\protect\epsilon _{+}$, evaluated
by the formula from ref. \protect\cite{ParazB20}, as a function of $\protect%
\epsilon _{+}/\protect\omega $ \ for $2\protect\pi u_{0}/a_{1}=0$ (dashed
curve), $2\protect\pi u_{0}/a_{1}=6.07$ (full curve), and as function of $2%
\protect\pi u_{0}/a_{1}$ (right panel) for the positron energy corresponding
to $\protect\epsilon _{+}/\protect\omega =0.5$. The values for the other
parameters are as follows: $\protect\psi =0.00552,$ $\protect\omega =20GeV,$
$\protect\nu _{s}=5\cdot 10^{9}$ Hz.}
\label{e+e-20gev}
\end{figure}

Numerical calculation show, that in dependence of the values for parameters,
the external excitation can either enhance or reduce the cross-section of
the pair creation process. As an illustration of the enhancement in the
cross-section integrated over the angle $\theta _{+}$, on the left panel of
Fig. \ref{e+e-20gev} we have plotted the quantity $10^{-3}(m_{e}^{2}\omega
/e^{6})d\sigma _{\pm }^{c}/d\epsilon _{+}$, evaluated by using the formula
from ref. \cite{ParazB20}, as a function of the ratio $\epsilon _{+}/\omega $
\ in the case of ${\mathrm{SiO}}_{2}$ monocrystal and Moliere
parameterization of the screened atomic potential for $2\pi u_{0}/a_{1}=0$
(dashed curve), $2\pi u_{0}/a_{1}=6.07$ (full curve). On the right panel the
same quantity is plotted as a function of $2\pi u_{0}/a_{1}$ for the
positron energy corresponding to $\epsilon _{+}/\omega =0.5$. The values for
the other parameters are taken as follows: $\omega =20$ GeV, $\psi =0.00552$%
, $\nu _{s}=5\cdot 10^{9}$ Hz for the frequency of acoustic waves. For the
amplitude of the deformation field corresponding to the numerical data of
Fig. \ref{e+e-20gev} the relative displacement of the neighboring atoms is
of the order $10^{-3}\mathring{A}$, which is much smaller than the
interatomic distance ($\sim 5\mathring{A}$).

\begin{figure}[tbph]
\begin{center}
\begin{tabular}{c}
\epsfig{figure=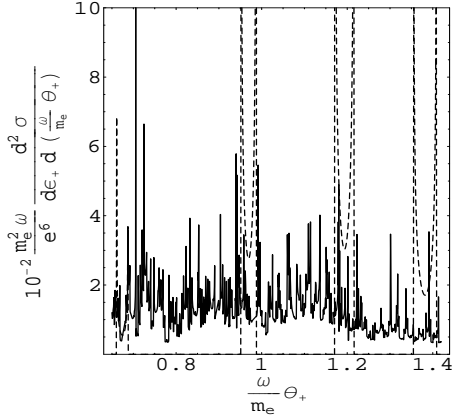,width=6cm,height=6cm}%
\end{tabular}%
\end{center}
\caption{Coherent pair creation cross-section, $10^{-3}(m_{e}^{2}\protect%
\omega /e^{6})d^{2}\protect\sigma _{\pm }^{c}/d\protect\epsilon _{+}d\left(
\protect\omega \protect\theta _{+}/m_{e}\right) $, evaluated by formula (
\protect\ref{sigsumg127}), as a function of $\protect\omega \protect\theta %
_{+}/m_{e}$ for $2\protect\pi u_{0}/a_{1}=0$ (dashed curve), $2\protect\pi %
u_{0}/a_{1}=6.07$ (full curve), $\protect\psi =0.00552$ . The values for the
other parameters are as follows: $\protect\epsilon _{+}/\protect\omega =0.5,$
$\protect\omega =20GeV,$ $\protect\nu _{s}=5\cdot 10^{9}$ Hz for the
frequency of acoustic waves.}
\label{e+20gevdis}
\end{figure}

For these values of parameters, when one has an enhancement of the
cross-section integrated over the angle $\theta _{+}$, we have numerically
analyzed the angular dependence of the pair creation cross-section by making
use of formula (\ref{sigsumg127}). In Fig. \ref{e+20gevdis} the quantity $%
10^{-3}(m_{e}^{2}\omega /e^{2})d^{2}\sigma _{\pm }^{c}/d\epsilon _{+}d\theta
_{+}$ is depicted as a function of $\omega \theta _{+}/m_{e}$ in the case of
${\mathrm{SiO}}_{2}$ monocrystal for $u_{0}=0$ (dashed curve) and $2\pi
u_{0}/a_{1}=6.07$ (full curve). The values for the other parameters are
taken as follows: $\epsilon _{+}/\omega =0.5,$ $\omega =20$ GeV,$\ \nu
_{s}=5\cdot 10^{9}$ Hz, $\psi =0.00552$.

\begin{figure}[tbph]
\begin{center}
\begin{tabular}{ccc}
\epsfig{figure=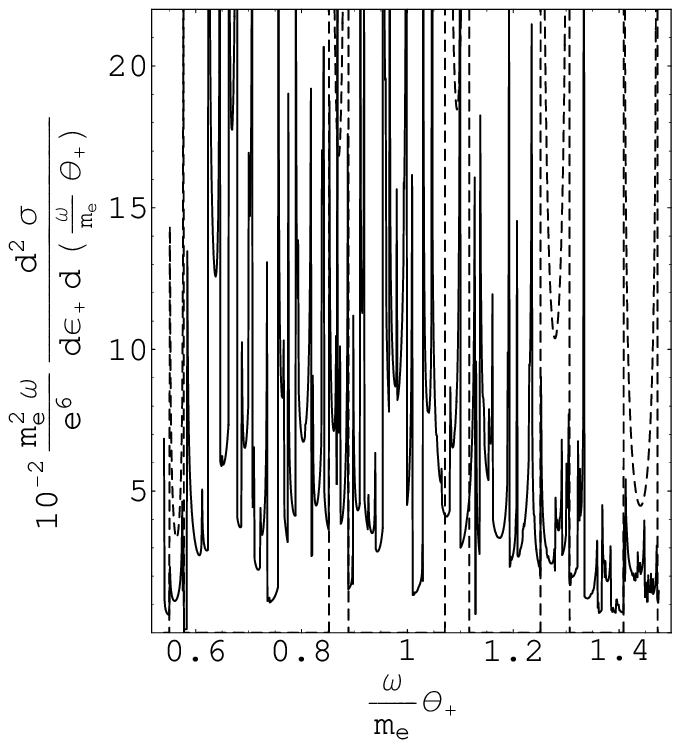,width=6cm,height=6cm} & \hspace*{0.5cm} & %
\epsfig{figure=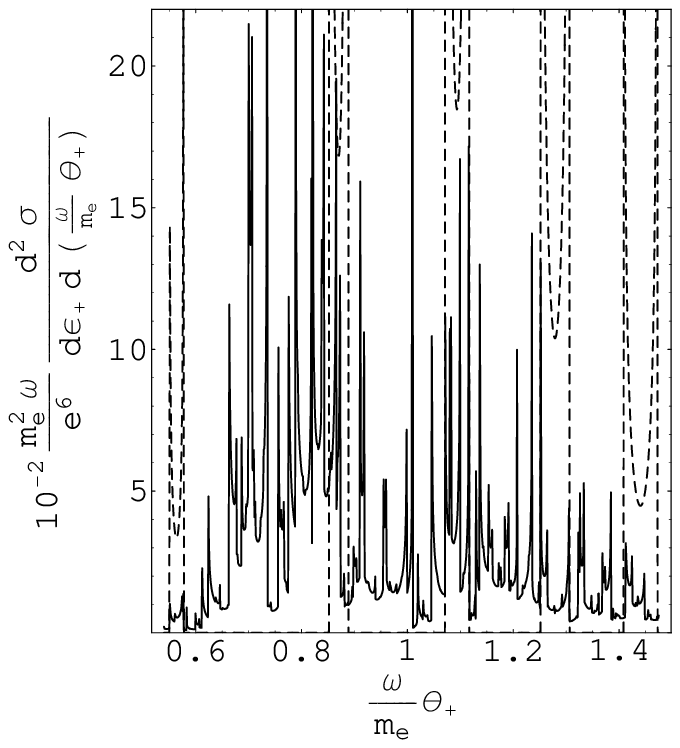,width=6cm,height=6cm}%
\end{tabular}%
\end{center}
\caption{Coherent pair creation cross-section for ${\mathrm{SiO}}_{2}$
crystal, $10^{-3}(m_{e}^{2}\protect\omega /e^{6})d^{2}\protect\sigma _{\pm
}^{c}/d\protect\epsilon _{+}d\left( \protect\omega \protect\theta %
_{+}/m_{e}\right) $, evaluated by formula ( \protect\ref{sigsumg127}), as a
function of $\protect\omega \protect\theta _{+}/m_{e}$ for $2\protect\pi %
u_{0}/a_{1}=0$ (dashed curve), $2\protect\pi u_{0}/a_{1}=1.1$ (full curve), $%
\protect\psi =0.001$ (left panel) and as a function of $\protect\omega
\protect\theta _{+}/m_{e}$ for $2\protect\pi u_{0}/a_{1}=0$ (dashed curve), $%
2\protect\pi u_{0}/a_{1}=2.14$ (full curve), $\protect\psi =0.001$ (right
panel). The values for the other parameters are as follows: $\protect%
\epsilon _{+}/\protect\omega =0.5,$ $\protect\omega =100GeV,$ $\protect\nu %
_{s}=5\cdot 10^{9}$ Hz for the frequency of acoustic waves.}
\label{quartz100}
\end{figure}

In order to see the dependence of the results on the energy of the incoming
photon, in Fig. \ref{quartz100} we have presented the quantity $%
10^{-3}(m_{e}^{2}\omega /e^{2})d^{2}\sigma _{\pm }^{c}/d\epsilon _{+}d\theta
_{+}$ as a function of $\omega \theta _{+}/m_{e}$ in the case of ${\mathrm{%
SiO}}_{2}$ monocrystal for the values of parameters $\epsilon _{+},$ $\psi $%
, $u_{0}$ taken from ref. \cite{ParazB20}, for which the integrated
cross-section is enhanced (reduced) by the acoustic wave. The dashed curves
on both panels correspond to the situation when the deformation field is
absent ($u_{0}=0$). The full curve on the left (right) panel is for the
amplitude of the deformation field corresponding to the value $2\pi
u_{0}/a_{1}=1.1$ (left panel, enhanced) ($2\pi u_{0}/a_{1}=2.14$, right
panel, reduced). The values for the other parameters are as follows: $%
\epsilon _{+}/\omega =0.5$, $\omega =100$ GeV, $\nu _{s}=5\cdot 10^{9}$ Hz, $%
\psi =0.001$.

\begin{figure}[tbph]
\begin{center}
\begin{tabular}{ccc}
\epsfig{figure=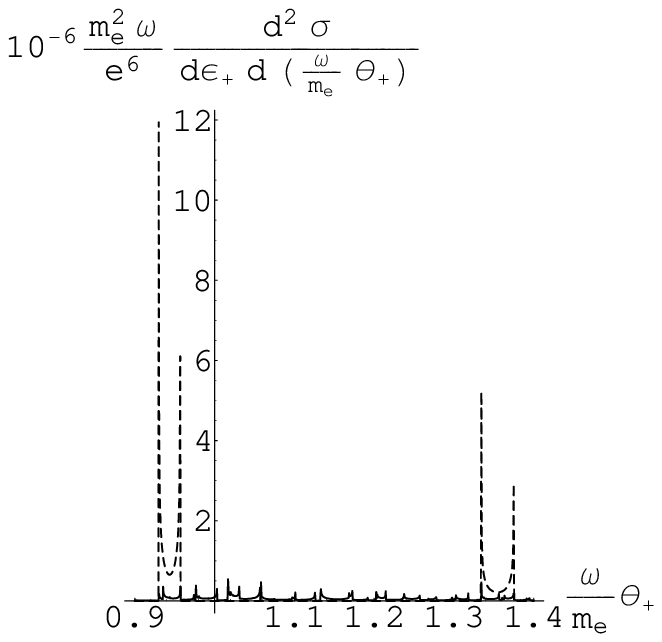,width=6cm,height=6cm} & \hspace*{0.5cm}
& \epsfig{figure=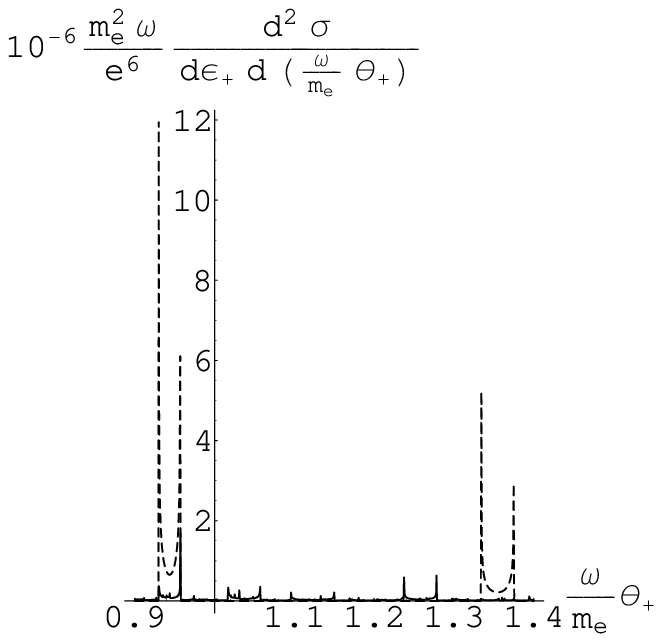,width=6cm,height=6cm}%
\end{tabular}%
\end{center}
\caption{Coherent pair creation cross-section for diamond crystal, $%
10^{-6}(m_{e}^{2}\protect\omega /e^{6})d^{2}\protect\sigma _{\pm }^{c}/d%
\protect\epsilon _{+}d\left( \protect\omega \protect\theta _{+}/m_{e}\right)
$, evaluated by formula ( \protect\ref{sigsumg127}), as a function of $%
\protect\omega \protect\theta _{+}/m_{e}$ for $2\protect\pi u_{0}/a_{1}=0$
(dashed curve), $2\protect\pi u_{0}/a_{1}=2.5$ (full curve), $\protect\psi %
=0.00142$ (left panel) and as a function of $\protect\omega \protect\theta %
_{+}/m_{e}$ for $2\protect\pi u_{0}/a_{1}=0$ (dashed curve), $2\protect\pi %
u_{0}/a_{1}=3.8$ (full curve), $\protect\psi =0.00142$ (right panel). The
values for the other parameters are as follows: $\protect\epsilon _{+}/%
\protect\omega =0.5,$ $\protect\omega =100GeV,$ $\protect\nu _{s}=5\cdot
10^{9}$ Hz for the frequency of acoustic waves.}
\label{diam100}
\end{figure}

It is also interesting to see the dependence of the results presented before
on the type of crystal. In Fig. \ref{diam100} we have plotted the quantity $%
10^{-6}(m_{e}^{2}\omega /e^{2})d^{2}\sigma _{\pm }^{c}/d\epsilon _{+}d\theta
_{+}$ as a function of $\omega \theta _{+}/m_{e}$ in the case of diamond
monocrystal for $u_{0}=0$ (dashed curves), $2\pi u_{0}/a_{1}=2.5$ (left
panel, full curve, enhanced) and for $2\pi u_{0}/a_{1}=3.8$ (right panel,
full curve, reduced). The values for the other parameters are taken as
follows: $\epsilon _{+}/\omega =0.5,$ $\omega =100$ GeV, $\nu _{s}=5\cdot
10^{9}$ Hz, $\psi =0.00142$.

As we see from the presented examples, the presence of the deformation field
leads to the appearance of additional peaks in the angular distribution of
the emitted positron (or electron) even for such ranges of values of an
angle of positron momentum, where due to the requirement $-1\leq y\leq 1$
the cross-section of process is zero when the deformation is absent. As we
have already mentioned before, this is related to that in the presence of
the deformation field the number of possibilities to satisfy the condition $%
g_{m\parallel }\geq \delta _{\pm }+g_{m\perp }^{2}/\left( 2\omega \right) $
in the summation in formula (\ref{sigcohgeneral}) increases due to the
presence of the additional term $mk_{s\parallel }$ in the expression for $%
g_{m\parallel }$.

\section{Conclusion}

\label{sec4:conc}

The present paper is devoted to the investigation of the angular
distribution of the positron in the pair creation process by high-energy
photons in a crystal with a complex lattice base in the presence of
deformation field of an arbitrary periodic profile. The latter can be
induced, for example, by acoustic waves. The influence of the deformation
field can serve as a possible mechanism to control the angular-energetic
characteristics of the created particles. The importance of this is
motivated by that the basic source to creating positrons for high-energy
colliders is the electron-positron pair creation by high-energy photons. In
a crystal the cross-section is a sum of coherent and incoherent parts. The
coherent part of the cross-section per single atom, averaged on thermal
fluctuations, is given by formula (\ref{sigcohgeneral}). In this formula the
factor $\left\vert F_{m}\left( \mathbf{g}_{m}\mathbf{u}_{0}\right)
\right\vert ^{2}$ is determined by the function describing the displacement
of the atoms due to the deformation field, and the factor $\left\vert
S\left( \mathbf{g}_{m},\mathbf{g}\right) \right\vert ^{2}$ is determined by
the structure of the crystal elementary cell. Compared with the
cross-section in an undeformed crystal, formula (17) contains an additional
summation over the reciprocal lattice vector of the one-dimensional
superlattice induced by the deformation field. We have argued that the
influence of the deformation field on the cross-section can be remarkable
under the condition $4\pi ^{2}u_{0}/a\gtrsim \lambda _{s}/l_{c}$. Note that
for the deformation with $4\pi ^{2}u_{0}/a>1$ this condition is less
restrictive than the naively expected one $\lambda _{s}\leq l_{c}$. The role
of coherence effects in the pair creation cross-section is essential when
the photon enters into the crystal at small angles with respect to a
crystallographic axis. In this case the main contribution into the coherent
part of the cross-section comes from the crystallographic planes, parallel
to the chosen axis (axis $z$ in our consideration). The behavior of this
cross-section as a function on the positron energy essentially depends on
the angle $\theta $ between the projection of the photon momentum on the
plane $\left( x,y\right) $ and $y$-axis. When the photon enters into the
crystal near a crystallographic plane, two cases have to be distinguished.
For the first one $\theta \sim a_{2}/2\pi l_{c}$ the formula (\ref%
{sigcasegz0}) is further simplified to the form (\ref{sigsumgxgy26}) under
the assumption $\mathbf{u}_{0}\perp \mathbf{a}_{1}$. In the second case one
has $\psi =\alpha \theta \sim a_{1}/2\pi l_{c}$, and the main contribution
into the cross-section comes from the crystallographic planes parallel to
the incidence plane. The corresponding formula for the cross-section takes
the form (\ref{sigsumg127}). The numerical calculations for the
cross-section are carried out in the case of ${\mathrm{SiO}}_{2}$ single
crystal with the Moliere parametrization of the screened atomic potentials
and for the deformation field generated by the transversal acoustic wave of $%
S$ - type with frequency 5 GHz. In order to illustrate the dependence of the
results on the type of crystal we have also presented the results for the
diamond monocrystal. Examples of numerical results are presented in figures.
The numerical calculations for values of the parameters in the problem when
one has an enhancement of the cross-section show that, the presence of the
deformation field leads to the appearance of additional peaks in the angular
distribution of the radiated positron (or electron) even for such ranges of
values of an angle of a positron, where due to the requirement $-1\leq y\leq
1$ the cross-section is zero when the deformation is absent. This can be
used to control the parameters of the positron sources for storage rings and
colliders.

\section*{Acknowledgment}

I am grateful to Aram Saharian for valuable discussions and suggestions.

\end{document}